\documentclass[prb,aps,twocolumn,showpacs,nobibnotes,epsf]{revtex4}
\usepackage{graphicx}% Include figure files
\usepackage{dcolumn}% Align table columns on decimal point
\usepackage{bm}% bold math
\usepackage{SIunits}

\begin{document}
\title{Oxygen Isotope Effect on the Spin State Transition in (Pr$_{0.7}$Sm$_{0.3}$)$_{0.7}$Ca$_{0.3}$CoO${_3}$}
\author{G. Y. Wang, T. Wu, X. G. Luo, W. Wang}
\author{X. H. Chen}
\altaffiliation{Corresponding author} \email{chenxh@ustc.edu.cn}
\affiliation{Hefei National Laboratory for Physical Science at
Microscale and Department of Physics, University of Science and
Technology of China, Hefei, Anhui 230026, People's Republic of
China\\}

\begin{abstract}
Oxygen isotope substitution is performed in the perovskite cobalt
oxide (Pr$_{0.7}$Sm$_{0.3}$)$_{0.7}$Ca$_{0.3}$CoO${_3}$ which
shows a sharp spin state transition from the intermediate spin
(IS) state to the low spin (LS) state at a certain temperature.
The transition temperature of the spin state up-shifts with the
substitution of $^{16}O$ by $^{18}$O from the resistivity and
magnetic susceptibility measurements. The up-shift value is 6.8 K
and an oxygen isotope exponent ($\alpha_S$) is about -0.8. The
large oxygen isotope effect indicates strong electron-phonon
coupling in this material. The substitution of $^{16}$O by
$^{18}$O leads to a decrease in the frequency of phonon and an
increase in the effective mass of electron ($m$$^\ast$), so that
the bandwidth W is decreased and the energy difference between the
different spin states is increased. This is the reason why the
$T_s$ is shifted to high temperature with oxygen isotopic
exchange.

\end{abstract}


\vskip 300 pt

\maketitle

\section{INTRODUCTION}

Perovskite-related cobalt oxides have attracted much interest
recently because of the spin state transition induced by
temperature. In a cubic crystal field, the 5-fold 3$d$ orbitals
will be split into two parts: 2-fold $e$$_g$ orbitals and 3-fold
$t$$_{2g}$ orbitals. The Co$^{3+}$ ion, which contains six 3$d$
electrons, often exibits a spin state transition from the low
spin(LS, $t$$_{2g}$$^6$) ground state to the intermediate spin(IS,
$t$$_{2g}$$^5$$e$$_g$$^1$) or to the high spin(HS,
$t$$_{2g}$$^4$$e$$_g$$^2$) state with increasing
temperature.\cite{Senaris,Asai,Kobayashi,Martin,Maignan,Moritomo,Saito,Vogt,Soda,Tsubouchi1,Tsubouchi2,Fujita1}
This indicates that the energy difference $\delta$$E$ between
these spin states is rather small, and the crystal field splitting
energy $\sl{\triangle}$$_C$ and the Hund coupling energy $J$$_H$
are comparable.

In the study of R$_{1-x}$A$_x$CoO$_3$ (R = Y and variousrare earth
elements, A = Ba, Sr, and Ca), Tsubouchi $et$
 $al$.\cite{Tsubouchi1,Tsubouchi2} found an abrupt metal-insulator (M-I) transition
and a spin state transition from IS state to LS state at 90 K with
decreasing temperature in Pr$_{0.5}$Ca$_{0.5}$CoO$_3$, which is
accompanied with a decrease of Co-O-Co bond angle. The temperature
at which the spin state transition occurs is defined as $T$$_S$.
They believe that the decrease of Co-O-Co bond angle results in a
reduction of covalency and thus unstablizes the itinerant IS
state. The volume contraction induced by the decrease Co-O-Co bond
angle can enlarge the splitting of crystal field, which also
stablize the LS state. Fujita $et$
 $al$.\cite{Fujita1} studied the $T$-$x$ phase diagram of Pr$_{1-x}$Ca$_{x}$CoO$_3$ under various
external pressures and found that this transition is not an
order-disorder transition. They also studied other
materials\cite{Masuda,Fujita1,Fujita2}, but only the cobalt oxides
which contain Pr and Ca show this transition. Based on the $x$
dependence of unit cell volume $V$$_u$, CoO$_6$ octahedra volume
$V$$_o$ and the Co-O-Co bond angle, they found that the Co-Co
transfer energy $t$ is also important to determine $\delta$$E$ in
this kind of material\cite{Fujita1}. But it is still unclear that
why only the materials contain Pr and Ca show this transition. In
order to give more experimental evidence for this question, we
investigate the oxygen isotope effect of
(Pr$_{0.7}$Sm$_{0.3}$)$_{0.7}$Ca$_{0.3}$CoO${_3}$ with a spin
state transition\cite{Fujita2}. The reason we choose this material
is that the oxygen content of this material is not so sensitive to
the heat treatment comparing to
Pr$_{0.5}$Ca$_{0.5}$CoO$_3$\cite{Fujita1}.

Isotope substitution is an effective tool to study the lattice
dynamical and electronic properties of solids\cite{Plekhanov}. In
perovskite maganites, a study on the effect of substitution of
$^{16}$O with $^{18}$O in La$_{0.8}$Ca$_{0.2}$MnO$_{3}$ shows a
large decrease in Curie temperature\cite{zhao1}. This result
significantly exceeds the temperature shifts of magnetic and
electronic phase transitions in the other earlier studied oxides
such as high-temperature superconductors, suggesting a strong
coupling of charge carriers to the Jahn-Teller lattice distortions
in manganites. This experiment also shows us an evidence for the
existence of polaron in this material. Moreover, a M-I transition
induced by oxygen isotope substitution was reported for
(La$_{0.5}$Nd$_{0.5}$)$_{0.67}$Ca$_{0.33}$MnO$_3$\cite{zhao2}and
(La$_{0.25}$Pr$_{0.75}$)$_{0.7}$Ca$_{0.3}$MnO$_3$\cite{Babushkina1,Babushkina2}.
 In cobalt oxides, electron-phonon coupling is
also important to affect the transport and magnetic
properties\cite{Senaris,Asai,Kobayashi,Soda,Tsubouchi1,Tsubouchi2,Fujita1,Fujita2,Masuda}.
For example, the Co$^{3+}$ in IS state is also a Jahn-Teller
active ion just like the Mn$^{3+}$ ion. Therefore, the interesting
phenomenon ought to be expected when $^{16}$O is substituted by
$^{18}$O. In the present work, we find an up-shift of $T_S$ about
6.8 K in (Pr$_{0.7}$Sm$_{0.3}$)$_{0.7}$Ca$_{0.3}$CoO${_3}$, and an
isotope exponent about -0.8 is obtained. To our best knowledge,
this is the first report of oxygen isotope effect study carried on
cobalt oxides. The result is discussed in the frame work of
electron-phonon coupling.

\section{EXPERIMENT DETAILS}
Polycrystal samples of
(Pr$_{0.7}$Sm$_{0.3}$)$_{0.7}$Ca$_{0.3}$CoO${_3}$ were prepared by
conventional solid-state reaction. Proper molar ratios of
Pr$_6$O$_{11}$, Sm$_2$O$_3$, CaCO$_3$ and Co$_3$O$_4$ were mixed,
well ground and then calcined at 1000 $\celsius$ and 1100
$\celsius$ for 24h with intermittent grinding. The powder was
pressed into pellets and sintered at 1200 $\celsius$ in the
flowing oxygen for 24 h. We call the samples obtained here the
as-synthesized samples.

One pellet of the as-synthesized samples was picked out and cut
into two pieces for oxygen-isotope diffusion. Each piece was put
into an alumina boat which was sealed in a quartz tube filled with
1 bar oxygen (one for $^{16}$O and another for $^{18}$O) mounted
in a furnace. The quartz tubes formed parts of two identical
closed loops. The quartz tubes were heated at 1000 $\celsius$ for
24h and then slowly cooled to room temperature. No impurity phase
was found in powder X-ray diffraction patterns. The oxygen-isotope
enrichment is determined from the weight changes of both $^{16}$O
and $^{18}$O samples. An  electronic microbalancer
(Mitteler-Telado XL26) with 1 $\mu$g precision was used for sample
weighting. The $^{18}$O sample has $70\%(\pm1\%)$ $^{18}$O and
$30\%(\pm1\%)$ $^{16}$O. To make sure the isotope exchange effect,
back-exchange of $^{18}$O sample by $^{16}$O was carried out in
the same way and the weight change recorded showed a complete
back-exchange. The resistivities and Magnetic susceptibilities of
all the samples were measured. Resistivity measurements were
performed on a AC resistance bridge (Linear Research Inc Model
LR700) by the standard four-probe method. Magnetic susceptibility
measurements were performed in  a superconducting quantum
interference device (SQUID) magnetometer (MPMS-7XL, Quantum
Design) under the magnetic field of 1000 Oe.

\section{RESULT AND DISCUSSION}

\begin{figure}[ht]
%\captionstyle{flushleft} \onelinecaptionsfalse
\centering
\includegraphics[width=9cm]{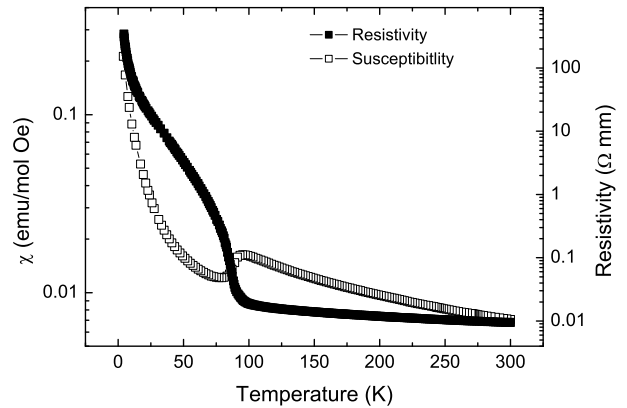}
\caption{The temperature dependence of the magnetic
susceptibilities and resistance of as-synthesized
(Pr$_{0.7}$Sm$_{0.3}$)$_{0.7}$Ca$_{0.3}$CoO${_3}$ sample. The spin
state transition can be seen in both curves.} \label{fig1}
\end{figure}

Figure 1 shows the temperature dependence of resistivity and
magentic susceptibility for the as-synthesized
(Pr$_{0.7}$Sm$_{0.3}$)$_{0.7}$Ca$_{0.3}$CoO${_3}$ sample. A sharp
decrease in susceptibility curve is found with decreasing
temperature, indicating the spin state change of Co$^{3+}$ from IS
to LS state\cite{Fujita2}. Here we define $T$$_S$ as the
temperature at which the susceptibility shows a maximum: 96 K for
(Pr$_{0.7}$Sm$_{0.3}$)$_{0.7}$Ca$_{0.3}$CoO${_3}$. A sharp
increase appears in the resistivity curve at $T$$_S$, which is
consistent with those reported by other group\cite{Fujita2}.
Co$^{3+}$ in IS state has one $e$$_g$ electron which can hop from
one site to another site via the Co$^{3+}$-O$^{2-}$-Co$^{4+}$
double exchange mechanism, and the $e$$_g$ electron here can be
regard as itinerant. Co$^{3+}$ in LS state has no $e$$_g$ electron
and its $t$$_{2g}$ orbitals are fully occupied, thus the charge is
localized. Therefore, its resistivity may jump several orders when
the Co$^{3+}$ ion abruptly change from IS to LS state.

\begin{figure}[ht]
%\captionstyle{flushleft} \onelinecaptionsfalse
\centering
\includegraphics[width=9cm]{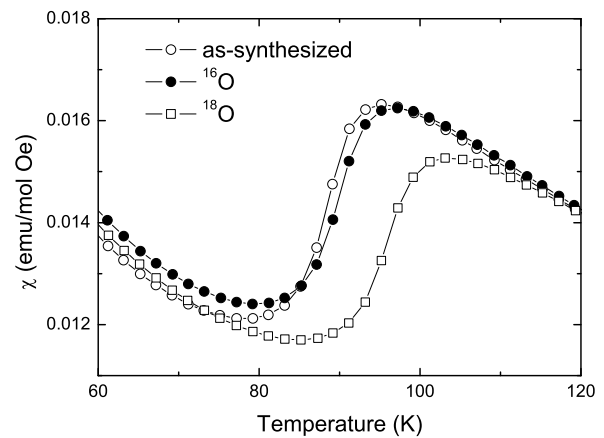}
\caption{The temperature dependence of magnetic susceptibility for
different samples. Open square for as-synthesized sample, closed
circle for $^{16}$O sample and open circle for $^{18}$O sample. A
positive shift of $T$$_S$ for $^{18}$O sample can be seen in this
figure.} \label{fig2}
\end{figure}

Figure 2 shows the magnetic susceptibilities of as-synthesized
sample, $^{16}$O sample and $^{18}$O sample for
(Pr$_{0.7}$Sm$_{0.3}$)$_{0.7}$Ca$_{0.3}$CoO${_3}$. A sharp
decrease in susceptibility curve is also found for $^{16}$O sample
and $^{18}$O sample. A very slight difference of $T$$_S$ between
as-synthesized sample and $^{16}$O sample is observed, this could
arise from the slight change of oxygen content or oxygen
distributing induced by 1000 $\celsius$ heat treatment. In
addition, the very slight change in $T_s$ suggests that the oxygen
content is robust in this system. The $^{18}$O and $^{16}$O
samples should have the same oxygen content due to the same heat
treatment in the same oxygen pressure. The $T_S$ we obtained: 97.1
K for the $^{16}$O sample and 103.9 K for the $^{18}$O sample. An
up-shift of $T$$_S$ is about 6.8 K with the substitution of
$^{16}$O by $^{18}$O. The oxygen isotope exponent $\alpha$$_S$ = -
dln$T$$_S$/dln$M$$_O$ (where $M$$_O$ is the oxygen isotope mass)
is about -0.8. This is a negative value just like the the exponent
for charge-ordered temperature in
La$_{0.57}$Ca$_{0.43}$MnO$_3$\cite{Isaac}. It is a remarkable
value compared to the reports. In La$_{1-x}$Ca$_x$MnO$_3$, the
exponent for Curie temperature $T$$_C$ is:
$\alpha$$_c$$\approx$0.85 for x=0.2 and 0.7 for x=0.1\cite{zhao1},
$\alpha$$_c$$\approx$0.57 for x=0.1, 1 for x=0.2, 0.55 for x=0.25,
0.32 for x=0.33 and 0.28 for x=0.4\cite{Zhao3}, respectively;
while the exponent for charge-ordered temperature is
$\alpha$$_{co}$$\approx$ -0.41 for x=0.43\cite{Isaac}. In
La$_{1-x}$Sr$_x$MnO$_3$, the exponent for Curie temperature
$T$$_C$: $\alpha$$_c$$\approx$0.19 for x=0.10, 0.14 for x=0.15 and
0.07 for x=0.3, repsectively\cite{zhao1}. In other materials, a
crossover from metal to insulating
phase\cite{zhao2,Babushkina1,Babushkina2,Unjong} induced by the
substitution of oxygen isotope are reported. The large oxygen
isotope exponent indicates strong electron-phonon coupling in this
material. The resistivity curves of as-synthesized sample,
$^{16}$O and $^{18}$O sample for
(Pr$_{0.7}$Sm$_{0.3}$)$_{0.7}$Ca$_{0.3}$CoO${_3}$ are shown in
Fig. 3. All the samples show a similar behavior that a M-I
transition occurs with decreasing temperature. An up-shift of the
M-I transition temperature by about 7 K is observed in $^{18}$O
sample, being consistent with the magnetic susceptibility results.

To ensure the oxygen isotope effect, a back-exchange of $^{18}$O
sample by $^{16}$O was performed. The resistivity data are shown
in Fig. 3. Its resistivity almost completely return to the
$^{16}$O sample, and its $T_s$ is the same as the $^{16}$O sample.
This result indicates that the oxygen isotope effect in
(Pr$_{0.7}$Sm$_{0.3}$)$_{0.7}$Ca$_{0.3}$CoO${_3}$ is intrinsic. It
 further confirms our explanation for the different $T$$_S$ between
as-synthesized sample and $^{16}$O sample.

\begin{figure}[t]
%\captionstyle{flushleft} \onelinecaptionsfalse
\centering
\includegraphics[width=9cm]{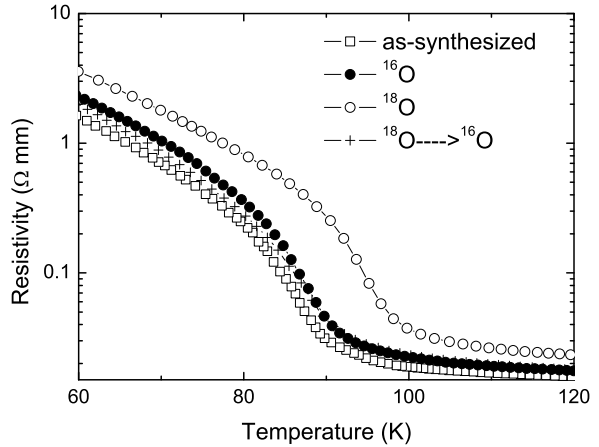}
\caption{Temperature dependence of resistivity for different
samples. Open square for as-synthesized sample, closed circle for
$^{16}$O sample, open circle for $^{18}$O sample and the plus
symbol for the back-exchange sample. A positive shift of M-I
transition temperature for $^{18}$O sample can also be seen in
this figure.} \label{fig3}
\end{figure}

The substitution of $^{16}$O with $^{18}$O leads to an unusual
large change in $T_s$ determined by the magnetic and transport
properties in (Pr$_{0.7}$Sm$_{0.3}$)$_{0.7}$Ca$_{0.3}$CoO${_3}$.
We have checked the X-ray diffraction patterns of $^{16}$O and
$^{18}$O samples, no obvious difference is found and almost same
lattice parameters are obtained. Therefore, the oxygen isotope
exchange does not change $\triangle$$_C$ and $J$$_H$, which
directly affect $\delta$$E$. The change of spin state transition
caused by the substitution of $^{16}$O with $^{18}$O should arise
from the phonon behavior. It suggests that there must be strong
electron-phonon coupling in this system since large change in
resistivity and magnetic susceptibility is induced. The
substitution of $^{16}$O by $^{18}$O reduces the frequency of
phonon, and then enhances the effective mass of electron
($m$$^\ast$) through the strong electron-phonon coupling. In
tight-binding approximation, the bandwith W is proportional to
1/$m$$^\ast$. Therefore, the enhancement of $m$$^\ast$ leads to a
decrease of W. The slight increase in high-temperature resistivity
for $^{18}$O sample compared to the $^{16}$O sample also indicates
a decrease of W. In other words, a decrease in kinetic energy of
electrons. When $\triangle_C$ and $J_H$ remain unchange, the
decrease of W leads to an increase $\delta$$E$, and consequently
causes an up-shift of $T$$_S$. On the other hand, Fujita $et$
$al$. have proved that $t$ is important to determine
$\delta$$E$\cite{Fujita1,Fujita2}. When $t$ decreases, $\delta$$E$
increases and the LS state will be stabilized. In Hubbard model, W
is proportional to 2z$t$, z is the coordinate number. Therefore,
the decrease of W leads to a decrease of $t$, then the increase of
$T$$_S$. Our result brings forward a direct evidence to support
the opinion of Fujita $et$ $al$..  Based on the observation of the
M-I transition induced by oxygen isotope exchange in
(La$_{0.5}$Nd$_{0.5}$)$_{0.67}$Ca$_{0.33}$MnO$_3$\cite{zhao2}and
 (La$_{0.25}$Pr$_{0.75}$)$_{0.7}$Ca$_{0.3}$MnO$_3$\cite{Babushkina1,Babushkina2}
 whose compositions are in the boundary between metallic and
 insulating phases in the phase diagram,
an IS-LS spin state transition induced by oxygen isotope effect
should be expected in a special $y$ for
(Pr$_{1-y}$Sm$_y$)$_{0.7}$Ca$_{0.3}$CoO$_3$ whose low temperature
phase changes from metal to insulator with increasing
$y$\cite{Fujita2}. This work is under investigation.

\section{CONCLUSION}

An up-shift by about 6.8 K in $T_S$ is observed in
(Pr$_{0.7}$Sm$_{0.3}$)$_{0.7}$Ca$_{0.3}$CoO${_3}$ when $^{16}$O is
substituted by $^{18}$O. The isotope exponent ($\alpha_S$) is
about -0.8. A strong electron-phonon coupling is responsible for
this large oxygen isotope effect. The effective mass of electron
is enhanced by the substitution, leading to a decrease in the
bandwidth and an increase in $\delta E$. Therefore, the increase
in $\delta$$E$ results in the up-shift of $T$$_S$.

\section*{ACKNOWLEDGMENTS} This work is supported by the grant from
the Nature Science Foundation of China and by the Ministry of
Science
and Technology of China, the Knowledge Innovation Project of Chinese Academy of Sciences.\\

\end{document}